# Partial-Wave Series Expansion and Angular Spectrum Decomposition Formalisms for Acoustical Beams

F.G. Mitri*

*Abstract* – Complex weights factors (CWFs) that fully define the incident beam independent of the presence of a scatterer, may be represented mathematically by either a partial-wave series expansion (PWSE) of multipoles, or the method of angular spectrum decomposition (ASD) of plane waves. Once the mathematical expression for the CWFs of the incident waves is known, evaluation of the arbitrary scattering, radiation force, and torque components on an object in 3D, using the Generalized Theories of Resonance Scattering (GTRS), Radiation Force (GTRF) and Radiation Torque (GTRT), becomes feasible. The aim of this Letter is to establish the connection between these two approaches in the framework of the GTRS, GTRF and GTRT in spherical coordinates for various acoustical applications. The advantage of using the ASD approach is also discussed for specific beams with particular properties.

PREDICTIONS and numerical computations of the arbitrary scattering, radiation force and torque components for a particle (or multiple particles) placed randomly in the field of finite (bounded) beams are important topics, playing significant roles in particle manipulation, mixing and rotation, medical, sonar and nondestructive imaging, therapeutic ultrasound and underwater acoustics to name a few applications.

The available analytical formalisms in acoustic scattering, which are used to evaluate the radiation force and torque components in 3D, are based on either the partial-wave series expansion (PWSE) method, in which the incident acoustical beam is modeled using a weighted PWSE [1-3], or a "hybrid" method [4], known as the angular spectrum decomposition (ASD) [5-7], in which a weighted sum of individual (monochromatic) plane waves propagating in different directions is used, and subsequently expanded into a PWSE using the scalar Laplace spherical harmonics functions. In both approaches, either the weighting function (WF) of the plane waves in the ASD, or the beam-shape coefficients (BSCs) in the PWSE, have to be determined to allow for the evaluation of the scattering, radiation force and torque components. Assuming that the incident beam expression is an exact solution of Helmholtz equation, both approaches should be equivalent and commensurate with the same result. Though the connection has been explored from the standpoint of optical diffraction theory [8], it is important to extend the analysis to its acoustic counterpart, and place it in the framework of the GTRS [2, 3], GTRF [9-12] and GTRT [13]. The aim here is to establish the connection between the PWSE and ASD formalisms by relating the BSCs (in PWSE) to the WF of the plane waves (in ASD). It is important to relate both formalisms so that applications involving the GTRS, GTRF and GTRT will strongly benefit from this relationship.

Consider an acoustical wave field emerging from an arbitrarily shaped aperture, and propagating in a nonviscous fluid. The incident complex pressure field, satisfies the Helmholtz equation,

$$\left(\nabla^2 + k^2\right) p_i = 0, \qquad (1)$$

where $k$ is the wave number. In a system of spherical coordinates $(r, \theta, \phi)$, the nonsingular solution of the Helmholtz equation is given by a PWSE as (Ch. 6 in [7]),

$$p_i^{PWSE}(\mathbf{r}) = \sum_{n=0}^{\infty} \sum_{m=-n}^{n} a_{nm} j_n(kr) Y_n^m(\theta, \phi), \qquad (2)$$

where $j_n(.)$ is the spherical Bessel function of the first kind, $Y_n^m(.)$ are the Laplace spherical harmonics functions, and $a_{nm}$ are defined as the BSCs in the PWSE method. The time-dependence factor $\exp(-i\omega t)$ is suppressed from (2) since the space-dependent pressure is only considered. Integrating (2) over a sphere of arbitrary radius $r$ using the orthogonality condition of the Laplace spherical harmonics [1], the expression for the BSCs is given by,

$$a_{nm} = \frac{1}{j_n(kr)} \int_{\phi=0}^{2\pi} \int_{\theta=0}^{\pi} p_i^{PWSE}(\mathbf{r}) Y_n^{m*}(\theta, \phi) \sin\theta \, d\theta \, d\phi, \qquad (3)$$

where the superscript * denotes a complex conjugate. The incident field expressed by (2) constitutes the key description of the incident beam in terms of the PWSE method.

Alternatively, an elementary solution of (1) is the plane wave, expressed mathematically by $\exp(i\mathbf{k}\cdot\mathbf{r})$, where $\mathbf{k} = (k_x, k_y, k_z)$ is the wave vector, and $\mathbf{r} = (x,y,z)$, is the vector coordinate of a point in space with respect to the origin. Accordingly, the pressure field constituting an exact solution of (1), is written as a weighted sum of plane waves in a Fourier transform space as (See (12) in [4]),

$$p_i^{ASD}(\mathbf{r}) = \frac{1}{4\pi^2} \int_{-\infty}^{+\infty}\int_{-\infty}^{+\infty} \Psi(k_x, k_y) e^{i\mathbf{k}\cdot\mathbf{r}} dk_x dk_y, \qquad (4)$$

where $\Psi(k_x, k_y)$ is a Fourier amplitude function, known as the angular spectrum, which characterizes the amplitude of each individual plane wave. The integration region in (4) is chosen so as to neglect evanescent waves (that decay exponentially away from the source, thus, do not contribute to the propagated field for $z > 0$) such that $k_x^2 + k_y^2 \leq k^2$. Note that the angular spectrum function is determined from the initial pressure $p_i^{ASD}(\mathbf{r})$ at $z = 0$, such that,

$$\Psi(k_x, k_y) = \int_{-\infty}^{+\infty}\int_{-\infty}^{+\infty} p_i^{ASD}(x, y, 0) e^{-i(k_x x + k_y y)} dx \, dy. \qquad (5)$$

*Corresponding author: F.G. Mitri (e-mail: F.G.Mitri@ieee.org).

For an individual plane wave traveling with arbitrary polar and azimuthal angles $(\theta', \phi')$, the expansion of the exponential function in terms of the Laplace spherical harmonics (See (10.43) in [14]),

$$e^{i\mathbf{k}\cdot\mathbf{r}} = 4\pi \sum_{n=0}^{\infty} \sum_{m=-n}^{n} i^n j_n(kr) Y_n^{m*}(\theta', \phi') Y_n^m(\theta, \phi), \quad (6)$$

leads to the expression for the incident pressure field using the ASD approach after substituting (6) into (4) as,

$$p_i^{ASD}(\mathbf{r}) = \frac{1}{\pi} \sum_{n=0}^{\infty} \sum_{m=-n}^{n} i^n \psi_{nm} j_n(kr) Y_n^m(\theta, \phi), \quad (7)$$

where $\psi_{nm} = \int_{-\infty}^{+\infty} \int_{-\infty}^{+\infty} \Psi(k_x, k_y) Y_n^{m*}(\theta', \phi') dk_x dk_y$.

The Fourier integral $\psi_{nm}$ represents the WF of the plane waves in the ASD approach, which solely defines the characteristics of the incident beam.

Equating (2) with (7), the connection between the BSCs and the WF of the plane waves is now established, such that,

$$\boxed{a_{nm} = \frac{i^n}{\pi} \psi_{nm}.} \quad (8)$$

The relationship (8) provides the main contribution of this work, in that it institutes the association of the PWSE method with the ASD approach.

Substituting (8) into (3) in [2] (with $P_0 = 1$), the scattered pressure of any arbitrary finite (or infinite) beam incident upon a scatterer at an arbitrary angle, given previously by (35b) in [4], is recovered. Moreover, the substitution of (8) into the 3D components of the acoustic radiation force functions (which are the radiation force per unit energy density and unit cross-sectional surface of the sphere), given initially by (11)-(13) in [9] (or (1),(2) in [10] – where $a_1^m$ in (2) should have been printed as $a_l^m$) leads *exactly*, after some algebraic manipulation, to (46)-(48) in [15], calculated using the ASD. Furthermore, inserting (8) into (14) in [13], gives the 3D dimensionless torque components in the ASD approach.

It is also important to emphasize that when a closed-form mathematical expression for the incident (usually infinite) beam in Cartesian coordinates is known, the BSCs in (3) have been evaluated using time-efficient numerical integration procedures with quadratures based on a Riemann sum [1], or using the Discrete Spherical Harmonics Transform (DSHT) [2, 10, 13], stemming from the PWSE formalism. Moreover, for a finite beam for which a closed-form expression for the *axial* BSCs can be derived, such as the case of a circular piston transducer [16, 17], a finite zero-order Bessel beam [18, 19], or finite vortex-type beams [20], the translational addition theorem of spherical wave function has been efficiently applied in evaluating the arbitrary scattering [3] and radiation force components [11] for a sphere. Nevertheless, for a finite (or infinite) beam that may not be described by a closed-form expression in Cartesian coordinates, or when the axial BSCs cannot be defined using a PWSE (such as a finite elliptical, square or rectangular linear array source etc.), or even when the beam description does not satisfy the Helmholtz equation (HE), for example in the case of a Gaussian beam, the ASD approach is more advantageous over the PWSE method in evaluating the WF, and subsequently the BSCs [via (8)]. This is because a plane wave ASD can be accomplished using (5), which is difficult to build using the PWSE approach. Eq.(8) can be used to advantage in connecting both formalisms, and makes it possible to determine the arbitrary scattering, radiation force and torque components for cases where the PWSE may not be applied, and *vice versa*. For a beam description that is not an exact solution of the HE, the ASD provides a remodeling of the beam (pp. 48-50 in [21]), and makes it an exact solution of the HE. Thus, the ASD approach presents an advantage over the PWSE in evaluating the BSCs, particularly for beams that generally do not satisfy the HE.